\documentclass[aps,pre,english,showpacs,showkeys,preprintnumbers,onecolumn,floatfix,nofootinbib,12pt]{revtex4-2}
\usepackage{eurosym}
\usepackage{amssymb}
\usepackage{graphicx,color}
\usepackage{epsfig}
\usepackage{rotating}
\usepackage[T1]{fontenc}
\usepackage{latexsym,epsfig}
\usepackage[latin9]{inputenc}
\usepackage{amsfonts}
\usepackage{dcolumn}
\usepackage{bm}
\usepackage{babel}
\usepackage{hyperref}
\usepackage{amsmath,mathrsfs}

\input{tcilatex}
\begin{document}

\title{Identifying Patterns Using Cross-Correlation Random Matrices Derived
from Deterministic and Stochastic Differential Equations}
\author{Roberto da Silva, Sandra D. Prado}
\affiliation{Institute of Physics, Federal University of Rio Grande do Sul, Brazil}
\email{Corresponding author: rdasilva@if.ufrgs.br}

\begin{abstract}
\end{abstract}

\begin{abstract}
Cross-Correlation random matrices have emerged as a promising indicator of
phase transitions in spin systems. The core concept is that the evolution of
magnetization encapsulates thermodynamic information [R. da Silva, Int. J.
Mod. Phys. C, 2350061 (2023)], which is directly reflected in the
eigenvalues of these matrices. When these evolutions are analyzed in the
mean-field regime, an important question arises: Can the Langevin equation,
when translated into maps, perform the same function? Some studies suggest
that this method may also capture the chaotic behavior of certain systems.
In this work, we propose that the spectral properties of random matrices
constructed from maps derived from deterministic or stochastic differential
equations can indicate the critical or chaotic behavior of such systems. For
chaotic systems, we need only the evolution of iterated Hamiltonian
equations, and for spin systems, the Langevin maps obtained from mean-field
equations suffice, thus avoiding the need for Monte Carlo (MC) simulations
or other techniques.
\end{abstract}

\maketitle

\section{Introduction}

\label{Section:Introduction}

Random matrices theory (RMT) is pivotal in deciphering the intricate
behavior of quantum systems exhibiting chaotic dynamics. In the realm of
quantum chaos, the statistical properties of energy levels and wavefunctions
are central to understanding the underlying physics. Random matrices provide
a powerful mathematical framework for modeling these properties, revealing
universal patterns such as Wigner-Dyson distributions and level repulsion 
\cite{Mehta} --- hallmarks of quantum chaotic systems \cite{Bohigas,Haake}.
Although originating in Physics in the context of Nuclear Physics and
quantum mechanics with connections with statistical physics \cite%
{Wigner,Dyson}, random matrices theory has since found broad application
across various disciplines, including economics \cite%
{Stanley,Stanleyb,Stanley2,Bouchaud,Bouchaud2}, neuroscience \cite{Bansal},
and complex networks \cite{Metz1,Metz2}.

In chaotic systems, characterized by extreme sensitivity to initial
conditions, this theory offers a rigorous approach to analyzing the
statistical properties of complex behaviors, enriching our theoretical
understanding of quantum chaos and extending its relevance to fields where
complexity and disorder are paramount.

Additionally, the Langevin equation and random matrices theory are
intimately connected in the study of stochastic processes and complex
systems in physics. The Langevin equation, a fundamental tool for modeling
the dynamics of particles in fluctuating environments, describes the
evolution of systems under random forces. This stochastic differential
equation captures the interplay between deterministic and stochastic
components, making it a cornerstone in nonequilibrium statistical mechanics 
\cite{TaniaMario2014,VanKampen}.

When applied to the Langevin equation, random matrices theory provides deep
insights into the spectral properties of associated stochastic operators and
the distribution of eigenvalues (see for example \cite{Dyson2}). This
connection is particularly significant for understanding disordered systems
and the dynamics of large ensembles of interacting particles. By linking
these two powerful frameworks, researchers can explore universal statistical
patterns in complex systems, deepening our understanding of both classical
and quantum phenomena.

Systems with a discrete time variable are referred to as maps. Maps describe
the time evolution of significant quantities in physics, crucial for
analyzing the stability, criticality, and chaotic behavior of systems \cite%
{Lichtenberg}. For instance, given a Hamiltonian $\mathcal{H}(q,p,t)$, the
evolution of generalized positions and momenta is governed by Hamilton's
equations \cite{Goldstein}:

\begin{equation}
\begin{array}{ccc}
\frac{dq}{dt}=\frac{\partial \mathcal{H}}{\partial p} & \text{and} & \frac{dp%
}{dt}=-\frac{\partial \mathcal{H}}{\partial q}%
\end{array}%
\text{ }\ 
\end{equation}

The discretization of these differential equations results in deterministic
maps, which are generally expressed as: 
\begin{eqnarray}
p_{n+1} &=&p_{n}+g_{1}(q_{n})  \label{Eq:deterministic_evolution} \\
&&  \notag \\
q_{n+1} &=&q_{n}+g_{2}(p_{n+1})  \notag
\end{eqnarray}%
where $n=0,1,2...$ parametrizes time, and $g_{1}$ and $g_{2}$ are functions
specific to the problem under consideration.

Similarly, in the case of a spin system in the mean-field regime, the time
evolution of magnetization $m$ is generally described by the Langevin
equation \cite{Pleimling2,TaniaMario2014}: 
\begin{equation}
\frac{dm}{dt}=-c\ \frac{\delta F(m;K_{1},K_{2},...,K_{l})}{\delta m}+\Gamma
\xi (t)  \label{Eq:Langevin}
\end{equation}%
where $F(m;K_{1},K_{2},...,K_{l})$ is the free energy of the system, and $%
K_{1},K_{2},...,K_{l}$ are the system's parameters, which can include
temperature, field, and combinations of such quantities. Here, $N$
represents the system size, i.e., the number of spins, such that 
\begin{equation}
\Gamma \sim \frac{1}{\sqrt{N}}
\end{equation}%
which must vanish in the thermodynamic limit, i.e., $N\rightarrow \infty $,
while $\xi (t)$ represents white noise. Therefore: 
\begin{eqnarray}
\left\langle \xi (t)\right\rangle &=&0\text{ } \\
&&  \notag \\
\left\langle \xi (t)\xi (t^{\prime })\right\rangle &=&\delta (t-t^{\prime })%
\text{.}  \notag
\end{eqnarray}

For simplicity, we denote $\varphi (m|K_{1},K_{2},...,K_{l})=\frac{\delta
F(m,K_{1},K_{2},...,K_{l})}{\delta m}$. Using this, a map can be defined for
the time evolution of magnetization:

\begin{equation}
m_{n+1}=m_{n}-\varepsilon c\varphi (m_{n}|K_{1},K_{2},...,K_{l})+\varepsilon
\Gamma \xi _{n},  \label{Eq:Iterated_magnetization}
\end{equation}%
where $\varepsilon $ is a parameter arising from the time discretization of
the differential equation.

Thus, although both maps \ref{Eq:deterministic_evolution} and \ref%
{Eq:Iterated_magnetization} originate the dynamics of physical systems, they
embody different physical phenomena. However, the spectral properties of the
cross-correlation random matrices built from their time evolution can reveal
their underlying dynamics.

Considering the problem in the context of Monte Carlo (MC) simulations of
the Ising, Potts models \cite{RMT2023,RMT2023-2,RMT2023-3}, we have shown
that cross-correlation random matrices (CCRM), even with few time evolutions
of the magnetization in these spin systems and using Metropolis or Glauber
dynamics with a small number of MC steps, can accurately capture the
criticality of the model. Our conclusion was that the density of eigenvalues
exhibits gaps that govern the transition, which is reflected in the first
and second moments of this density.

It is noteworthy that several authors have explored the spectral properties
of CCRM in spin systems using different approaches from ours. For example,
Vinayak et al. \cite{Vinayak2014} investigated the spectral properties of
correlation matrices during near-equilibrium phase transitions. They studied
the correlation matrices of $N$ spins in the two-dimensional Ising model
across several MC steps, revealing evidence of power-law spatial
correlations at phase transitions. Similarly, Biswas et al. \cite{Biswas2017}
analyzed the steady-state correlation matrix in the asymmetric simple
exclusion process. It is important to note that dealing with matrices of
this size, proportional to the number of spins, is computationally
demanding. Our method, however, offers a significant advantage by operating
with much smaller matrices, as we will demonstrate.

Alternatively, maps from Hamiltonian systems can also be used to build CCRM
of one-parameter systems. In this direction, we showed that this method is
able to explore the chaoticity and nuances of the logistic map and the
Chirikov standard map by studying the spectra of these matrices \cite%
{RMT2023-4}.

In this current work, we take a step further. We propose to study the
spectral properties of CCRM from spin systems, not through MC simulations,
but by considering the proper Langevin equation describing their processes.
Specifically, we use the Ising model and its spin-1 version, the Blume-Capel
(BC) model. This approach is particularly interesting because it elevates
the method to the level of stochastic differential equations, rather than
relying on MC simulations.

Conversely, to corroborate that the properties of maps of different physical
systems can be translated by the spectra of random matrices, we studied the
Kicked Harper Map (KHM), a two-dimensional area preserving map that exhibits
mixed dynamics \cite{Artuso}. This map is the classical counterpart of the
quantum dynamical system known as the kicked-Harper model.

The work is organized as follows: In the next section, we explain how the
method is applied to the various systems discussed in this study. We then
present the results in Section \ref{Section:Results}, followed by summaries
and conclusions in the final section.

\section{Cross-correlation matrices of maps derived from deterministic and
stochastic differential equations}

Here, we define a cross-correlation matrix $\mathcal{C}$ by using time
series iterations of a given observable $O$ that can be a generalized moment 
$p$, generalized position $q$, magnetization per particle $m$, and so on.
This term, in the context of RMT, to the best of our knowledge, originated
in Econophysics \cite{Stanley,Stanley2,Stanleyb,Bouchaud,Bouchaud2} could
also be referred to as a covariance matrix or a correlation coefficient
matrix.

The elements of this matrix for a general observable $O$ can be defined as
follows:

\begin{equation*}
\mathcal{C}_{ij}\mathcal{=}\frac{\left\langle O^{(i)}O^{(j)}\right\rangle
-\left\langle O^{(i)}\right\rangle \left\langle O^{(j)}\right\rangle }{%
\sigma _{O^{(i)}}\sigma _{O^{(j)}}}
\end{equation*}%
which are calculated by sampling two time evolutions of length $N_{steps}$: $%
O_{0}^{(i)},O_{1}^{(i)},...,O_{n_{steps-1}}^{(i)}$ and $%
O_{0}^{(j)},O_{1}^{(j)},...,O_{n_{steps-1}}^{(j)}$, such that:%
\begin{equation*}
\left\langle O^{(k)}\right\rangle =\frac{1}{N_{steps}}%
\sum_{p=0}^{N_{steps-1}}O_{p}^{(k)}\text{, }\left\langle
O^{(i)}O^{(j)}\right\rangle =\frac{1}{N_{steps}}\sum_{p=0}^{N_{steps}-1}%
\sum_{q=0}^{N_{steps-1}}O_{p}^{(i)}O_{q}^{(j)}\text{ }
\end{equation*}%
with $\sigma _{O^{(k)}}=\frac{1}{(N_{steps}-1)}%
\sum_{p=0}^{N_{steps}-1}(O_{p}^{(k)}-\left\langle O^{(k)}\right\rangle
)^{2}\approx $ $\left\langle O^{(k)2}\right\rangle -\left\langle
O^{(k)}\right\rangle ^{2}$.

It is important to note that the matrix $\mathcal{C}$, of dimension $%
N_{sample}$, can be obtained from the standardized time-evolution matrix\ $%
\mathcal{M}$: 
\begin{equation}
\mathcal{M}=\left( 
\begin{array}{cccc}
\frac{O_{0}^{(1)}-\left\langle O^{(1)}\right\rangle }{\sigma _{O^{(1)}}} & 
\frac{O_{0}^{(2)}-\left\langle O^{(2)}\right\rangle }{\sigma _{O^{(2)}}} & 
\cdots & \frac{O_{0}^{(N_{sample})}-\left\langle
O^{(N_{sample})}\right\rangle }{\sigma _{O^{(N_{sample})}}} \\ 
\frac{O_{1}^{(1)}-\left\langle O^{(1)}\right\rangle }{\sigma _{O^{(1)}}} & 
\frac{O_{1}^{(2)}-\left\langle O^{(2)}\right\rangle }{\sigma _{O^{(2)}}} & 
& \frac{O_{1}^{(N_{sample})}-\left\langle O^{(N_{sample})}\right\rangle }{%
\sigma _{O^{(N_{sample})}}} \\ 
\vdots & \vdots &  & \vdots \\ 
\frac{O_{N_{steps}-1}^{(1)}-\left\langle O^{(1)}\right\rangle }{\sigma
_{O^{(1)}}} & \frac{O_{N_{steps}-1}^{(2)}-\left\langle O^{(2)}\right\rangle 
}{\sigma _{O^{(2)}}} &  & \frac{O_{N_{steps}-1}^{(N_{sample})}-\left\langle
O^{(N_{sample})}\right\rangle }{\sigma _{O^{(N_{sample})}}}%
\end{array}%
\right)
\end{equation}

This allows us to verify that: 
\begin{equation}
\mathcal{C=}\frac{1}{N_{steps}}\mathcal{M}^{t}\mathcal{M}.
\label{Eq:Covariance_matrix}
\end{equation}

It is known that if $O_{0}^{(i)},O_{1}^{(i)},...,O_{n_{steps-1}}^{(i)}$ are
a set of independent random variables, we are in the context of the real
Wishart ensemble \cite{Wishart,Seligman3,Novaes,Guhr}. In that case, the
joint probability distribution of the eigenvalues is expected to be given by:

\begin{equation}
P(\lambda _{1},...,\lambda _{N_{sample}})=Z^{-1}e^{-\mathcal{H}(\lambda
_{1},...,\lambda _{N_{sample}})}\text{,}
\end{equation}%
where $Z=\idotsint d\lambda _{1}...d\lambda _{N_{sample}}e^{-\mathcal{H}%
(\lambda _{1},...,\lambda _{N_{sample}})}$, and

\begin{equation*}
\mathcal{H}(\lambda _{1},...,\lambda _{N_{sample}})=\mathcal{H}%
_{auto}(\lambda _{1},...,\lambda _{N_{sample}})+\mathcal{H}_{int}(\lambda
_{1},...,\lambda _{N_{sample}})
\end{equation*}%
corresponds to a Hamiltonian of a Coulomb gas with logarithmic repulsion 
\begin{equation}
\mathcal{H}_{int}(\lambda _{1},...,\lambda _{N_{sample}})=-\sum_{i<j}\ln
\left\vert \lambda _{i}-\lambda _{j}\right\vert \text{,}  \label{Eq:int}
\end{equation}%
and the term 
\begin{equation}
\mathcal{H}_{auto}(\lambda _{1},...,\lambda _{N_{sample}})=\frac{N_{steps}}{2%
}\sum_{i=1}^{N_{sample}}\lambda _{i}-\frac{(N_{steps}-N_{sample}-1)}{2}%
\sum_{i=1}^{N_{sample}}\ln \lambda _{i}  \label{Eq:auto}
\end{equation}%
attracting the particles to the origin, with $\lambda _{1},...,\lambda
_{N_{sample}}\geq 0$. Since $N_{steps}>N_{sample}$, we have $\lambda >\frac{%
(N_{steps}-N_{sample}-1)}{N_{steps}}\ln \lambda $.

In the case of potentials described by Eqs. \ref{Eq:int} and \ref{Eq:auto},
the density of states defined by: 
\begin{equation}
\rho (\lambda )=\int_{0}^{\infty }\int_{0}^{\infty }...\int_{0}^{\infty
}P(\lambda ,\lambda _{2}...,\lambda _{N_{sample}})d\lambda _{2}d\lambda
_{3...}d\lambda _{N_{sample}}
\end{equation}
follows the well-known Marchenko-Pastur law \cite{Marcenko,Sengupta}:

\begin{equation}
\rho _{M}(\lambda )=\left\{ 
\begin{array}{l}
\dfrac{N_{MC}}{2\pi N_{sample}}\dfrac{\sqrt{(\lambda -\lambda _{-})(\lambda
_{+}-\lambda )}}{\lambda }\ \text{if\ }\lambda _{-}\leq \lambda \leq \lambda
_{+} \\ 
\\ 
0\ \ \ \ \text{otherwise,}%
\end{array}%
\right.  \label{Eq:MP}
\end{equation}%
where $\lambda _{\pm }=1+\frac{N_{sample}}{N_{MC}}\pm 2\sqrt{\frac{N_{sample}%
}{N_{MC}}}$.

As previously discussed, our observable $O$ can represent magnetization in
spin systems, or simply a moment or position in dynamical systems. An
interesting quantity to monitor, which describes how the density of states
deviates from the non-correlated situation given by $\rho _{M}(\lambda )$ in
Eq. \ref{Eq:MP}, is the first moment of $\rho (\lambda )$, i.e., $E[\lambda
]=\int_{0}^{\infty }\lambda \rho (\lambda )d\lambda $. Its estimator is
obtained using the numerical density of states:

\begin{equation}
\left\langle \lambda \right\rangle =\frac{\sum_{i=1}^{N_{int}}\rho
_{N}(\lambda _{i})\lambda _{i}}{\sum_{i=1}^{N_{int}}\rho _{N}(\lambda _{i})}%
\text{,}  \label{Eq:Average}
\end{equation}%
where $N_{int}$ is the number of bins used to calculate the numerical
density of states $\rho _{N}(\lambda _{i})$, which may differ from the
expression in Eq. \ref{Eq:MP}. In other words, the function $\mathcal{H}%
_{auto}(\lambda _{1},...,\lambda _{N_{sample}})$ in Eq. \ref{Eq:MP} changes,
leading to gaps in the density of states that are reflected in $\left\langle
\lambda \right\rangle $ or higher moments $\rho (\lambda )$. Our previous
work has shown that extreme values of $\left\langle \lambda \right\rangle $
are associated with phase transition critical points in spin systems \cite%
{RMT2023,RMT2023-2,RMT2023-3} and chaotic-to-stability transitions in
chaotic maps \cite{RMT2023-4}.

In this paper, we demonstrate that such transitions are intrinsically
present in the stochastic differential equations (specifically Langevin
equations) governing the dynamics of spin systems. We tested this using maps
from the Ising and Blume-Capel (BC) models.

As a second contribution, we show that this method can be extended to a
chaotic two-parameter dynamical system, the Kicked Harper model. In the next
section, we provide an overview of these systems and explain how the $%
\mathcal{C-}$matrices are applied using the Random Matrices Theory (RMT)
method described earlier.

\subsection{Maps Derived from Langevin Equations: Ising and BC Models}

In this case, we focus on Ising-like Hamiltonians, which can be generalized
as follows:

\begin{equation}
\mathcal{H}=-J\sum\limits_{\left\langle i,j\right\rangle }\sigma _{i}\sigma
_{j}+D\sum\limits_{i=1}^{N}\sigma _{i}^{2}-H\sum\limits_{i=1}^{N}\sigma _{j}%
\text{.}  \label{Eq:short_range_hamiltonian}
\end{equation}%
If $D=0$ and $\sigma _{j}=\pm 1$ (spin 1/2), we have the standard Ising
model. For $D\geq 0$ (anisotropy term) and $\sigma _{j}=0,\pm 1$ (spin 1),
we have the Blume-Capel model. Here, $H$ is the external field that couples
with each spin, and $\left\langle i,j\right\rangle $ denotes that the sum is
taken only over the nearest neighbors in a $d$-dimensional lattice.

A mean-field approximation assumes that $i$-th spin $\sigma _{i}$ interacts
with a magnetic "cloud" represented by the average magnetization $\xi _{i}=$ 
$\frac{1}{N}\sum_{j=1,j\neq i}^{N}\sigma _{j}$. Each spin, within its
original lattice, is linked to $z=2^{d}$ neighbors. The number of links in
the lattice is $Nz/2$, as each spin is counted with $z/2$ links to avoid
double-counting. Thus, in this approximation, the interaction term $%
H_{int}=-J\sum_{\left\langle i,j\right\rangle }\sigma _{i}\sigma _{j}$ must
be replaced in the mean-field approximation by:

\begin{equation}
\mathcal{H}_{int}^{(MF)}=-\frac{Jz}{2}\sum\limits_{i=1}^{N}\sigma _{i}\xi
_{i}\approx -\frac{Jz}{2N}\sum\limits_{i=1}^{N}\sum\limits_{j=1}^{N}\sigma
_{i}\sigma _{j}
\end{equation}

Finally, the mean-field Hamiltonian can be written as: 
\begin{equation}
\mathcal{H}^{(MF)}=-\frac{Jz}{2N}M^{2}-HM+D\sum\limits_{i=1}^{N}\sigma
_{i}^{2}\text{,}  \label{Eq:long_hange_hamiltonian}
\end{equation}%
where $M=\sum_{i=1}^{N}\sigma _{i}$ is the magnetization of the system. For
our purposes, we set $H=0$ from this point onward.

Focusing initially on the Ising model ($D=0$), the free energy of the system
is given by: 
\begin{equation}
\ f_{I}(m;K)=\frac{m^{2}}{2}-\frac{1}{K}\ln \left[ 2\cosh (Km)\right] \text{,%
}
\end{equation}%
where $K=\beta Jz$, with $\beta =\frac{1}{k_{B}T}$ and $z$ as the
coordination number. Thus, we can write the map for this case. Considering
the equation \ref{Eq:Iterated_magnetization} for this situation, we have:

\begin{equation}
m_{n+1}=(1-\varepsilon )m_{n}+\varepsilon \tanh (Km_{n})+\varepsilon \Gamma
\xi _{n}  \label{Eq:Iteration_ising}
\end{equation}

On the other hand, for the BC model, there are two parameters: $K_{1}=\beta
Jz$ and $K_{2}=\beta D$. Consequently, the free energy is:

\begin{equation}
f_{BC}(m;K_{1},K_{2})=\frac{m^{2}}{2}-\frac{1}{K_{1}}\ln (2e^{-K_{2}}\cosh
(K_{1}m)+1)
\end{equation}

The map in this case is somewhat more complicated: 
\begin{equation}
m_{n+1}=(1-\varepsilon )m_{n}+\frac{2\varepsilon e^{-K_{2}}\sinh (K_{1}m_{n})%
}{2e^{-K_{2}}\cosh (K_{1}m_{n})+1}+\varepsilon \Gamma \xi _{n}.
\label{Eq:Iteration_BC}
\end{equation}

\subsection{Map for a Two-Parameter Deterministic System: The Kicked-Harper
Model}

Chaotic behavior is fundamental in contemporary physics \cite{Lichtenberg},
as the understanding of non-deterministic phenomena under varying initial
conditions is pivotal in numerous contexts. This includes the stabilization
of seemingly simple mechanical systems, such as the inverted pendulum \cite%
{Peretti,Prado}, and investigations into the degenerate route to chaos in
dissipative systems \cite{Jason}.

The identification of chaos in specific Hamiltonian systems can be
accomplished using traditional methods; however, there is room for the
development of various alternatives. In contrast, the theory of random
matrices has provided a robust and potent toolkit for describing several
aspects of physical phenomena.

Area-preserving maps play an important role in diverse fields as they are
widely used for modeling complex systems. Maps of regular dynamics usually
present stable islands and chaotic behavior separately or together in the
phase space depending on the control parameter.

In this context, and interesting model is that of hamiltonian : 
\begin{equation}
\mathcal{H}(q,p,t)=L\cos p+K\cos q\sum_{n=-\infty }^{\infty }\delta (t-n)
\end{equation}%
corresponding to the pulsed version of Harper's Hamiltonian:

\begin{align}
\frac{dq}{dt}& =\frac{\partial \mathcal{H}}{\partial p}=-L\sin p \\
&  \notag \\
\frac{dp}{dt}& =-\frac{\partial \mathcal{H}}{\partial p}=K\sin
q\sum\limits_{n=-\infty }^{\infty }\delta (t-n)\   \notag
\end{align}

Hence, the Kicked Harper map, which preserves the area in the phase space of
the two canonical dynamical variables $q$ and $p$, is defined as follows:

\begin{align}
p_{n+1}& =p_{n}+K\sin q_{n}  \label{Eq:Kicked_Map} \\
q_{n+1}& =q_{n}-L\sin p_{n+1}  \notag
\end{align}

The dynamics can be visualized on a torus by taking $q\ $mod $2\pi $ and $p$
mod $2\pi $. Thus, the idea is to construct $\mathcal{C}$ using the time
evolutions of $p$ and $q$ as described by Eq. \ref{Eq:Kicked_Map}, and to
investigate the resulting structures with our RMT method.

\section{ Results}

\label{Section:Results}

We have separated our results into two parts. The first part is dedicated to
the study of maps derived from stochastic Langevin equations, which describe
the criticality of mean-field Ising and Blume-Capel models. In the second
part, we explore the applicability of our method in describing a map from a
deterministic system, specifically the two-parameter Kicked Harper model.

\subsection{Criticality of Spin Systems via the Langevin Equation}

We initially conducted iterations of the Ising map (Eq. \ref%
{Eq:Iteration_ising}), mirroring the approach employed in our previous study
using MC simulations \cite{RMT2023}. To achieve this, we consider a time
step of size $\varepsilon =10^{-4}$. In this case, we obtained our
observable represented by matrix elements $m_{i}^{(j)}$. Here, $%
i=0,...,N_{steps}-1$, where we used $N_{steps}=300$ iteration steps, and $j$
ranges from 1 to $N_{sample}=100$ different series.

To obtain the histograms $\rho _{N}(\lambda )\times \lambda $, we used a
total of $N_{run}=1000$ different matrices. For all histograms we considered 
$N_{bin}=100$ bins. This parameter set will also be used in all other
results explored in this paper. We chose $m_{0}^{(j)}$ $\in \lbrack -1,1]$
randomly.

An important numerical point, also reported in some of our previous works,
is related to constructing $\rho _{N}(\lambda )$ and calculating the average
according to equation \ref{Eq:Average}, rather than simply estimating it as
an arithmetic average: $\bar{\lambda}=N_{sample}^{-1}N_{run}^{-1}%
\sum_{i=1}^{N_{sample}}\sum_{j=1}^{N_{run}}\lambda _{ij}$. This approach is
crucial due to the sensitivity of the numerical results explored here.

Thus, we construct the ensemble of matrices $\mathcal{C}$ for the Ising
model, where the observables are captured by Eq. \ref{Eq:Iteration_ising}
according to Eq. \ref{Eq:Covariance_matrix}. The results for $\left\langle
\lambda \right\rangle $ as a function of $K$ are presented in Fig. \ref%
{Fig:Ising_model} for different values of $\Gamma $.

\begin{figure}[tbp]
\begin{center}
\includegraphics[width=1.0\columnwidth]{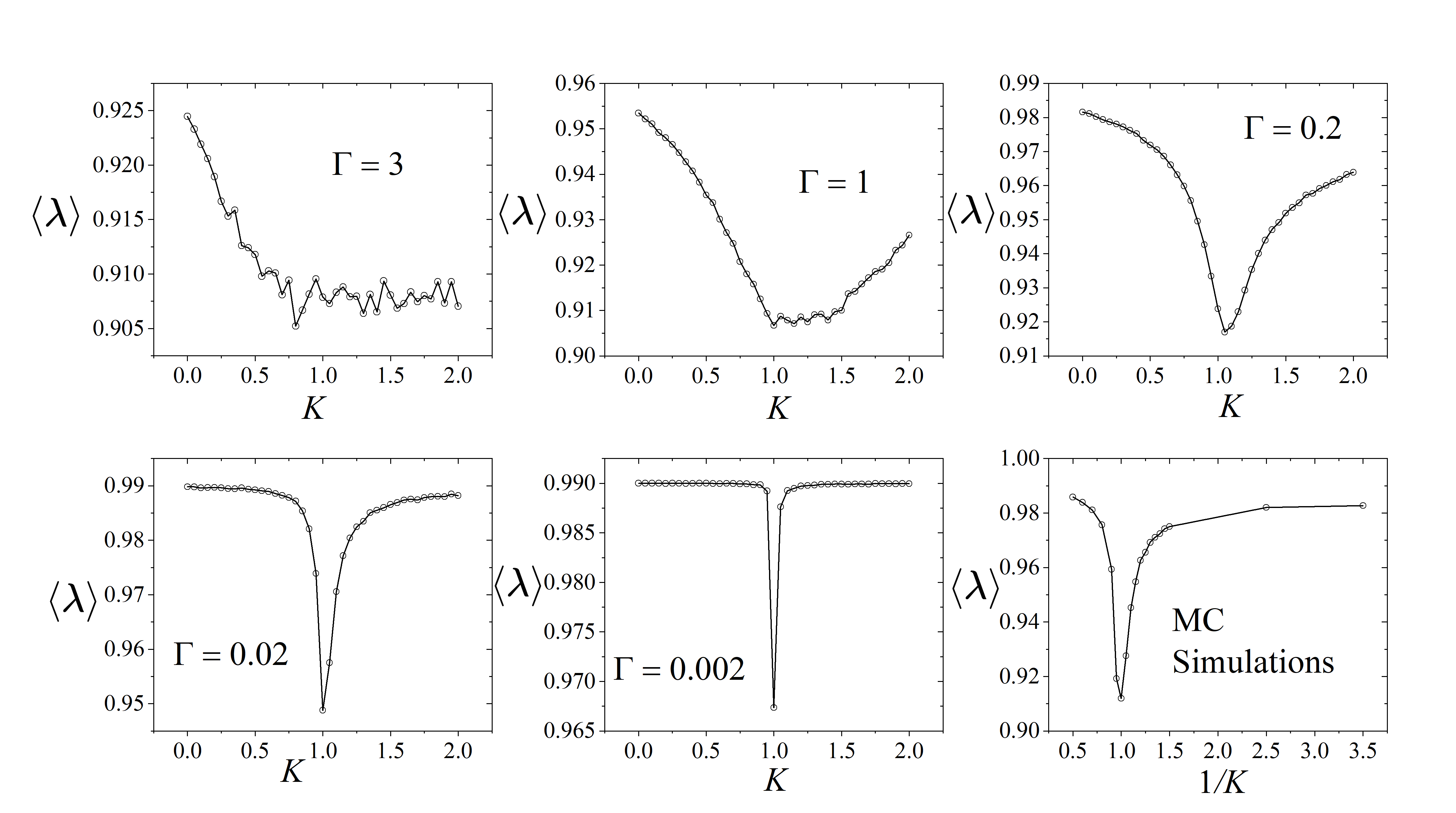}
\end{center}
\caption{The average eigenvalue of $\mathcal{C}$ as function of $K$ for
different values of $\Gamma $ is shown for the mean-field Ising map. The
expected transition at $K=1$ is observed in the thermodynamic limit ($\Gamma
\rightarrow 0$) and is corroborated by the method with MC simulations of
magnetization evolutions. The line connecting the points is not a fit; it is
provided solely for reference. }
\label{Fig:Ising_model}
\end{figure}

We observe that for small systems (large $\Gamma $), the average eigenvalue
does not accurately reflect the expected critical value of the model ($K=1$%
). However, as $\Gamma \rightarrow 0$ ($N\rightarrow \infty $), a minimum
value at $K=1$ is gradually approached. The final plot demonstrates that
this minimum value is confirmed by using time evolutions of magnetization
from MC simulations of the mean-field Ising model under the Metropolis
prescription (as shown in the last plot of Fig. \ref{Fig:Ising_model} with
data extracted from \cite{RMT2023-3}).

It is important to note that we expect $\rho (\lambda )$ to approach $\rho
_{M}(\lambda )$ (Eq. \ref{Eq:MP}) as $K\rightarrow 0$ ($T\rightarrow \infty $%
), which corresponds to the paramagnetic phase. In this situation, the $k-$%
th moment, $\left\langle \lambda ^{k}\right\rangle _{\text{numerical}}$,
should closely align with the theoretical value:

\begin{equation}
\left\langle \lambda ^{k}\right\rangle _{\text{exact}}=\int\limits_{-\infty
}^{\infty }\lambda ^{k}\rho _{M}(\lambda )d\lambda =\sum\limits_{j=0}^{k-1}%
\frac{\left( \frac{N_{sample}}{N_{MC}}\right) ^{j}}{j+1}\binom{k}{j}\binom{%
k-1}{j}\text{.}
\end{equation}%
where this relation is derived by expanding the binomials and applying the
well-known Vandermonde identity: $\sum_{l=0}^{r}\binom{m}{l}\binom{n}{r-l}=%
\binom{m+n}{r}$.

For $k=1$ (which is our focus), we expect $\left\langle \lambda
\right\rangle \approx 1$, as observed for $T\rightarrow \infty $ in Fig. \ref%
{Fig:Ising_model}. It is noteworthy that in these plots, $\left\langle
\lambda \right\rangle \approx 0.99$. This small discrepancy is due to the
fact that we are analyzing the evolution of magnetization rather than a
single spin, leading to some spurious correlations. The fact that $%
\left\langle \lambda \right\rangle \approx 1$ also for $T<T_{C}$ is not
related to $\rho (\lambda )$ approaching $\rho _{M}(\lambda )$; rather, it
is a coincidence or an idiosyncrasy of the method, as $\left\langle \lambda
\right\rangle \approx 1$ is a sufficient but not necessary condition. In
fact, $\rho (\lambda )$ differs significantly from $\rho _{M}(\lambda )$, as
observed in \cite{RMT2023}.

Thus, we can see that the method performs well at the differential equation
level for the Ising model, which motivates us to apply this methodology to
models with more complex phase diagrams, such as the Blume-Capel model.

The second-order phase transition line of the Blume-Capel (BC) model, in
mean-field aproximation, is described by the equation: 
\begin{equation}
K_{1}=1+\frac{e^{K_{2}}}{2}\text{.}  \label{Eq:theoretical_prediction}
\end{equation}

\begin{figure}[tbp]
\begin{center}
\includegraphics[width=1.0\columnwidth]{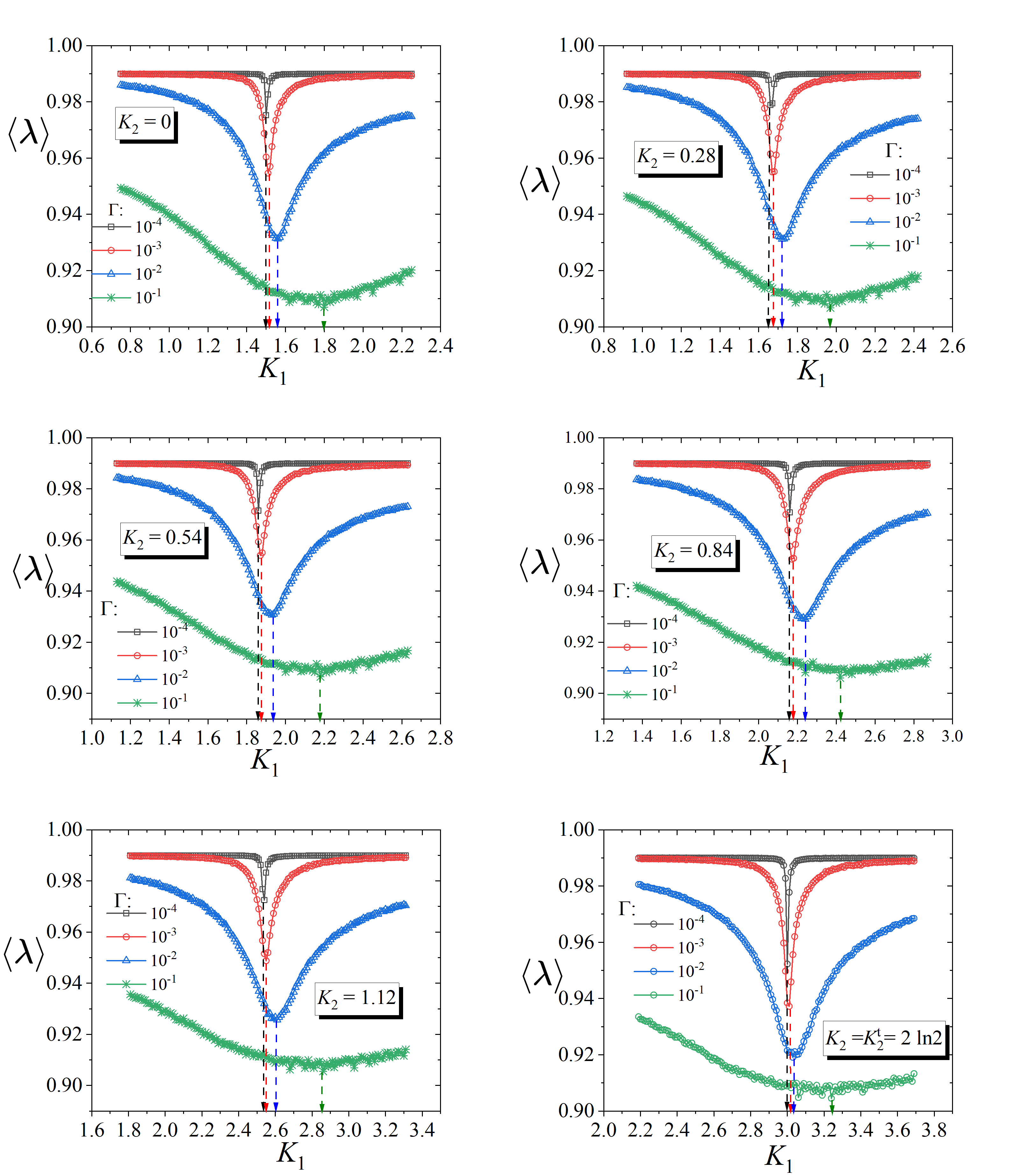}
\end{center}
\caption{Average eigenvalue as a function of $K_{1}$ for different values of 
$K_{2}$ along the critical line. }
\label{Fig:BC}
\end{figure}

This curve ends at $(K_{1},K_{2})=(3,\ln 4)$, which corresponds to the
tricritical point; beyond this point, the transition becomes first-order.
Thus, we iterate the equation \ref{Eq:Iteration_BC} by fixing $K_{2}$ and
varying $K_{1}$, while also observing the average eigenvalue, as shown in
Fig. \ref{Fig:BC}. Similarly, $m_{0}^{(j)}$ is randomly chosen from the
interval $[-1,1]$ and we also consider a time step of size $\varepsilon
=10^{-4}$.

\begin{figure}[tbp]
\begin{center}
\includegraphics[width=1.0\columnwidth]{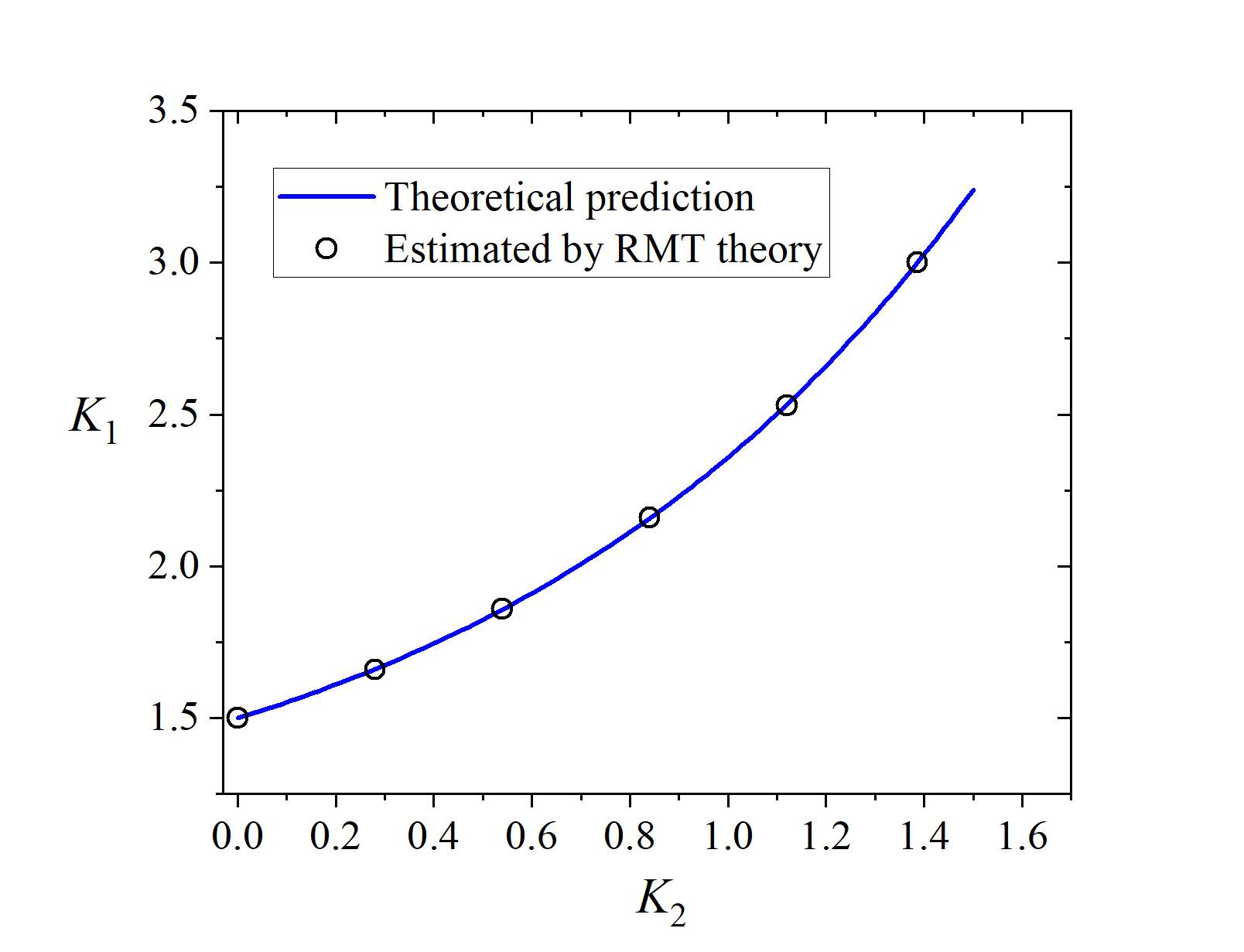}
\end{center}
\caption{Critical values estimated using the RMT method for the Blume-Capel
model align precisely with the theoretical curve.}
\label{Fig:BC_gran_finalle}
\end{figure}

Each plot considers systems of different sizes, and we observe that $%
K_{1}^{(C)}$ approaches the expected value as $N\rightarrow \infty \ $($%
\Gamma \rightarrow 0$). This is summarized in Fig. \ref{Fig:BC_gran_finalle}.

This figure demonstrates that the RMT method accurately identifies the
critical values. Notably, the use of Langevin equations to construct the
matrices appears to mitigate the crossover effects observed in the
simulations of the BC model (see \cite{EliseuRMT}).

\subsection{Characterization of the Chaotic Behavior in the Kicked Harper
Model}

Our algorithm constructs an ensemble of $N_{run}=1000$ distinct matrices $%
\mathcal{C}$ of dimension $N_{sample}\times N_{sample}$. These matrices
correspond to $N_{run}$ different initial conditions, which are randomly
selected with $q_{0},p_{0}\in \lbrack 0,2\pi ]$. We then diagonalize these
matrices and organize the eigenvalues within the range $\lambda _{\min }^{(%
\text{Numerical})}$ to $\lambda _{\max }^{(\text{Numerical})}$. Using a
fixed number of bins, $N_{bin}=100$, we generate histograms to calculate $%
\rho (\lambda _{i})$ in the same manner as we performed for the Ising \ and
BC model.

\begin{figure}[tbp]
\begin{center}
\includegraphics[width=1.0\columnwidth]{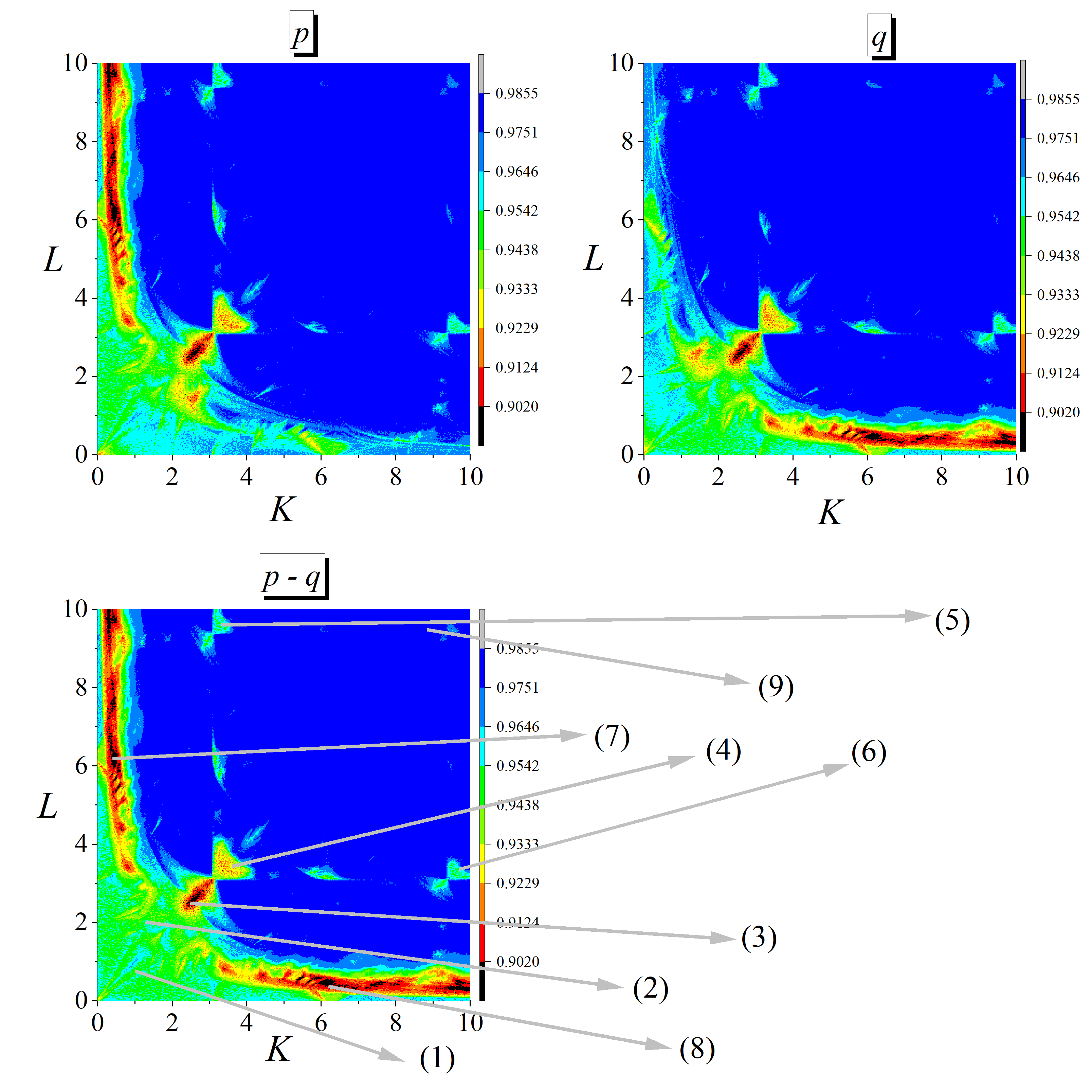}
\end{center}
\caption{Average eigenvalue is attributed for each $(K,L)$ parameter pair.
The first diagram shows the results for the eigenvalues using the evolution
of $p$, while the second uses $q$. A third diagram considers the upper part (%
$L>K$) with values of $p$, while a the lower part considers the values where
($L<K$). The arrows indicated in the plot highlight specific points for our
analysis. }
\label{Fig:map}
\end{figure}

Thus, we obtain the average eigenvalue $\left\langle \lambda \right\rangle $
for each pair of values $(K,L)$ as shown in Eq. \ref{Eq:Average}. The
results are presented in the first two diagrams: one considering the
evolution of $q$ and the other the evolution of $p$. Additionally, a third
diagram combines the information from the previous two diagrams, denoted as $%
p-q$. Fig. \ref{Fig:map} summarizes this information in a heat map.

The first diagram shows the results for the eigenvalues using the evolution
of $p$, and the second diagram uses $q$. The third diagram considers the
upper part $(L>K)$ with values of $p$ and the lower part $(L<K)$.

\begin{figure}[tbp]
\begin{center}
\includegraphics[width=1.0\columnwidth]{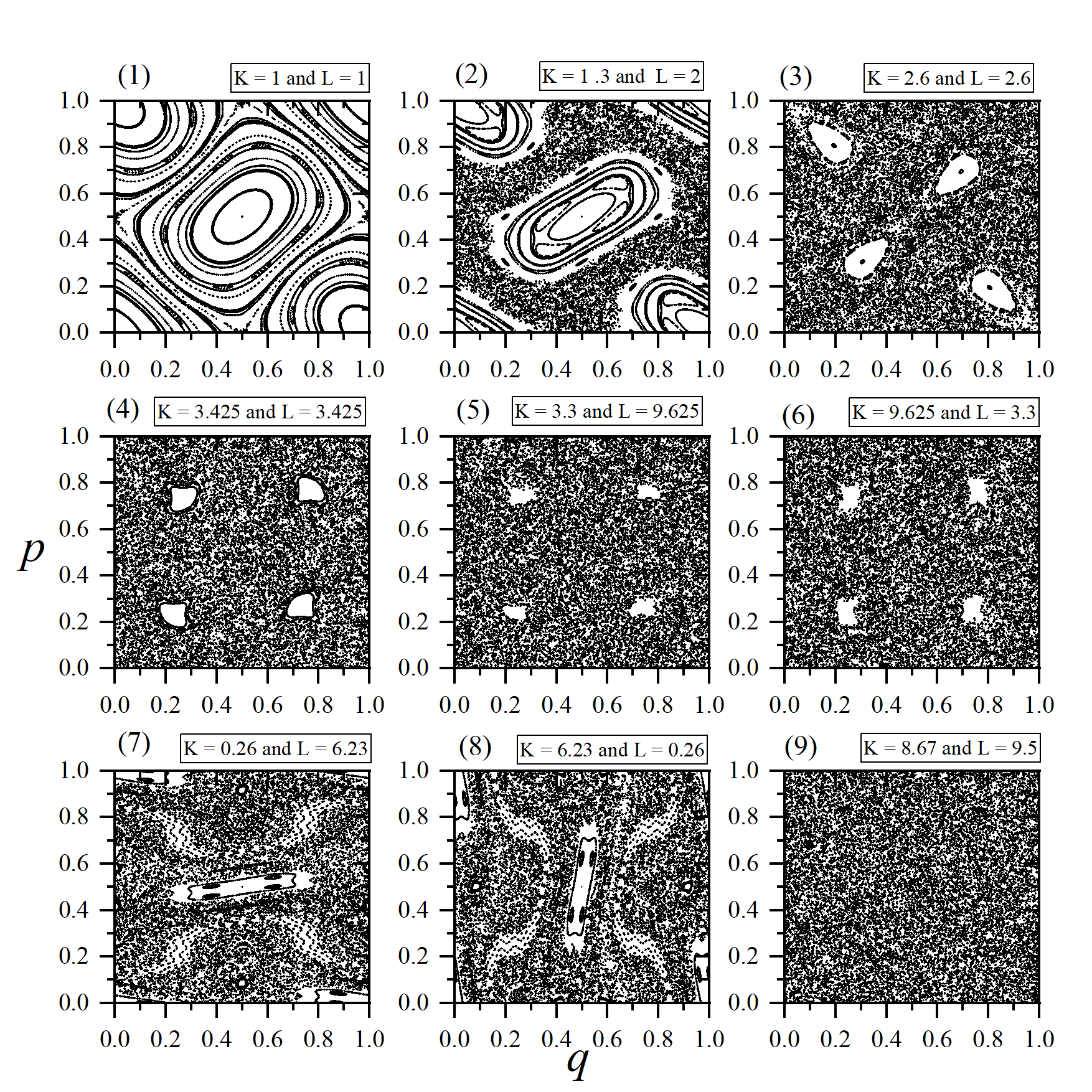}
\end{center}
\caption{Maps $p\times q$ are obtained from specific points selected in the
heat map shown in Fig. \protect\ref{Fig:map}. }
\label{Fig:phase_space}
\end{figure}

In this diagram, we observe a large dark blue region, which corresponds to
the area with a higher average value that we associate with chaos. Other
regions are painted in different colors, indicating various structures that
can be examined. In Fig. \ref{Fig:map}, we selected nine points in different
regions to observe the $p\times q$ maps corresponding to each of these
points. We begin with point 1 ($K=L=1$). This point lies in a green region
where yellow and green hues blend together. The corresponding phase space
for this point can be observed in Fig. \ref{Fig:phase_space}.

A stable structure emerges from this parameter choice. Investigating a
nearby point ($K=1.3$ and $L=2$) we also observe a stable structure with
small chaotic regions. This indefinition persists across different points
studied, as illustrated in this plot. Stable islands can be observed in the
"chaotic sea" for points indexed by 3 ($K=2.6$ and $L=2.6$) and 4 ($K=3.425$
and $K=3.425$) in the diagram, which are located in regions of low average
values (ranging from yellow to black).

We will now investigate some symmetrical points. Let us focus on points 5 ($%
K=3.3$ and $L=9.625$) and 6 ($K=9.625$ and $L=3.3$). It is interesting to
observe that both points exhibit patterns of holes in chaotic regions, but
unlike points 3 and 4, these holes do not have well-defined boundaries.
Additionally, the longitudinal direction of the holes is orthogonal when
swapping $K$ and $L$.

Next, we investigate another pair of symmetrical points: 7 ($K=0.26$ and $%
L=6.23$) and 8 ($K=6.23$ and $L=0.26$)). In these points, we observe
"insets" of stability within the chaotic regions. Finally, point 9 ($K=8.67$
and $L=9.5$) is located in the dark blue sea (chaotic region) \ref{Fig:map}.
The phase space in Fig. \ref{Fig:phase_space} accurately describes this
situation.

\section{Conclusions}

We demonstrate that a flexible Random Matrices Theory method, based on
Cross-correlation matrices, is highly effective for identifying phase
transitions in maps from spin systems evolved by the Langevin equation.
Similarly, the method is also very useful for describing chaotic behavior in
deterministic maps with two parameters. In the first case, the extremal
values of the average eigenvalue of such matrices indicate criticality,
while in the second case, smaller values of this parameter seem to indicate
some degree of stability. The method captures the nuances and patterns of
the systems at the differential equation level, thus avoiding the need for
extensive numerical simulations, which warrants further investigation.

\section*{Acknowledgments}

R. da Silva extends gratitude to the Aguia4 cluster at HPC-USP and the
Lovelace cluster at IF-UFRGS for providing the computational resources.
Additionally, thanks are due to CNPq for partial financial support of this
work under grant 304575/2022-4. The authors would like to thank F. L. Metz
for  highlighting an important point regarding the Langevin equation.

\end{document}